\documentclass[twocolumn,preprintnumbers,amsmath,amssymb]{revtex4}
\usepackage{dcolumn}
\usepackage{bm,epic,eepic}
\usepackage[english]{babel}

\newcommand{\av}[1]{\langle #1 \rangle}
\newcommand{\BEQ}{\begin{eqnarray}}
\newcommand{\EEQ}{\end{eqnarray}}
\newcommand{\ket}[1]{\left | \, #1 \right \rangle}

\newcommand{\bra}[1]{\left \langle \, #1 \right |}

\newcommand{\beq}{\begin{equation}}
\newcommand{\eeq}{\end{equation}}

\newcommand{\Eq}[1]{Eq.\,\,(\ref{#1})}

\newcommand{\forget}[1]{}

\begin{document}

\title{Separable quantum states do not
have stronger correlations than local realism.\\ A comment on quant-ph/0611126 of Z. Chen.}

\author{Michael Seevinck}
\email{seevinck@phys.uu.nl}
\affiliation{%
Institute for History and Foundations of Science,\\
 Utrecht University
 PO Box 80.000, 3508 TA Utrecht, the Netherlands}%

\date{\today}

\begin{abstract}
Chen (quant-ph/0611126) has recently claimed ``exponential violation of local realism by separable states", in the sense
that multi-partite separable quantum states are supposed to give rise to correlations and fluctuations that violate a Bell-type
inequality that Chen takes to be satisfied by local realism.
 However, this can not be true since all predictions (including all correlations and fluctuations)
 that separable quantum states give rise to have a local realistic description and thus satisfy all Bell-type inequalities,  
 and this holds for all number of parties.
Since Chen claims otherwise by presenting a new inequality, claimed to be a Bell-type one,
 which separable states supposedly can violate,
there must be a flaw in the argumentation. I will expose this flaw,
not merely for clarification of this issue, but perhaps even more importantly
 since it re-teaches us an old lesson
John Bell taught us over 40 years ago.
I will argue that this lesson provides us with
a new morale especially relevant to modern research in Bell-type inequalities.

\end{abstract}
\maketitle
\noindent
\begin{centerline}{\bf Introduction}\end{centerline}
\vskip0.4cm
\noindent
Chen \cite{chen} has recently claimed ``exponential violation of local realism by separable states", in the sense
that multi-partite separable quantum states are supposed to give rise to correlations and fluctuations that violate a Bell-type
inequality that Chen takes to be satisfied by local realism. The violation is claimed to be by an exponentially increasing
amount as the number of particles $n$ grows ($n>2$). In fact, it is claimed to violate the local realistic maximum by a factor of $2^{n-2}$.

 However, this can not be true since all predictions (including all correlations and fluctuations)
 that separable quantum states give rise to have a local realistic description and thus satisfy all Bell-type inequalities,
 and this holds for all number of parties.
Although this is known already, let me nevertheless show this first (for the sake of completeness of the discussion)
after which I will continue my comment on the work of Chen.  Here, I will generalise the exposition by
\.Zukowski \cite{zukowski} to the $n$-party case.

 Any $n$-party Bell-type inequality has the following generic form
\beq\label{bellineq}
|\sum_{k_1,k_2,\ldots,k_n} c(k_1,k_2,\ldots,k_n)E(k_1,k_2,\ldots,k_n)|\leq B(c),
\eeq
where $k_1$, $k_2$, etc. are labels that distinghuish various (discrete or continuous) measurement
settings that can be measured on system 1, 2, etc.,  $c(k_1,k_2,\ldots,k_n)$
are certain constant coefficients, $B(c)$ is the maximum value obtainable by a local realistic theory for
the expression on the left hand side of \Eq{bellineq},
and $E(k_1,k_2,\ldots,k_n)$ is the `correlation function' for outcomes of measurement with settings
$k_1,k_2,\ldots,k_n$, which for local realistic theories is assumed to be the expectation 
of the product of the local observables:
\begin{multline}
\label{lr}
E(k_1,k_2,\ldots,k_n)=\\\int_\Lambda  A_1(k_1,\lambda)A_2(k_2,\lambda)\ldots A_n(k_n,\lambda)\,p(\lambda)\,d \lambda.
\end{multline}
Here $\lambda$ is an integration variable, often denoted as the hidden variable,
the set $\Lambda$ is the set of hidden variables, and the functionals
$A_i(k_i,\lambda)$ are either the outcomes ($+1$ or $-1$) of the measurements determined by the settings $k_i$ and hidden variable $\lambda$
in case we are dealing with a deterministic theory, or they are
 the expectation values (in the interval $[-1,\,1]$) of these outcomes in the case 
 of a stochastic (non-deterministic) theory, 
 and finally $p(\lambda)$ is a normalised
probability distribution of the variable $\lambda$. The Bell inequalities
of \Eq{bellineq} follow solely from the assumption of \Eq{lr}.

 Let us now suppose that we have an $n$-partite fully
separable quantum state $\rho$, i.e., a convex sum of product states:
$\rho=\sum_j p_j \rho^1_j\otimes\rho^2_j\otimes\ldots\otimes\rho^n_j$, with
 $\rho^i_j$  a density operator for party (subsystem) $i$. 
The index $j$ is a summation or integration variable and $\sum_j p_j=1$, $0\leq p_j\leq 1$.
 The correlations these states give rise to all have the form
 \BEQ
&&\av{\hat{A}_1(k_1)\hat{A}_2(k_2)\hat{A}_n(k_n)}_\rho:= \nonumber\\ \label{corrqm}
&&\mathrm{Tr}[\rho\,\hat{A}_1(k_1)\otimes\hat{A}_2(k_2)\ldots \otimes\hat{A}_n(k_n)]=
\\
&& \sum_jp_j\mathrm{Tr}^1[\rho^1_j\hat{A}_1(k_1)]\mathrm{Tr}^2[\rho^2_j\hat{A}_2(k_2)]\ldots\mathrm{Tr}^n [ \rho^n_j\hat{A}_n(k_n)]\nonumber
,
\EEQ
with Tr$^i[\,\cdot\,]$ the partial trace for party $i$ (i.e., the rest of the parties is traced out),
and $\hat{A}_i(k_i)$ the operator associated with the measurement on party $i$ with setting $k_i$. 
Each such measurement has possible outcomes $+1$ or $-1$, 
just as was the case above \cite{hat}. Since we have that $|\mathrm{Tr}^i[\rho^i\hat{A}_i(k_i)]|=|\av{\hat{A}_i(k_i)}_\rho|\leq\max A_i(k_i,\lambda)$
the correlations of \Eq{corrqm} can be written as in \Eq{lr} and thus they can be reproduced by a local
realistic theory. They thus must satisfy all Bell-type inequalities of \Eq{bellineq},
i.e., `those known at present, as well as those which one day would be derived'\cite {zukowski}.
Note that the same holds for all fluctuations a separable state can give rise to, cf. \cite{fluct}.

Since Chen claims otherwise by presenting a new inequality, claimed to be a Bell-type one,
 which separable states supposedly can violate,
there must be a flaw in the argumentation. I will expose this flaw,
not merely for clarification of the issue, but perhaps even more importantly
 since it re-teaches us an old lesson
John Bell has made over 40 years ago, 
although in a different form. I will argue this lesson provides us with
a new morale especially relevant to modern research in Bell inequalities.

\vskip0.5cm
\begin{centerline}{\bf Review of Chen's results}\end{centerline}
\vskip0.4cm
\noindent
Let me first present the result of Chen's analysis. He considers the well-known
Mermin-Roy-Singh-Ardehali-Belinski\u{\i}-Klyshko inequality for $n$-parties 
\cite{mkrsabk}.
This inequality is characterised by a specific choise \cite{mermin} of 
coefficients $c(k_1,k_2,\ldots,k_n)$ 
in \Eq{bellineq}, where each party chooses between two observables 
(i.e., each $k_i$ has two possibilities) that are furthermore dichotomic ($\pm1$ valued). 
Chen takes the inequality to be normalised so that $B(c)=1$, and
chooses a specific choice  
of measurement settings $k_1,k_2\ldots,k_n$, such that the two possible 
local settings for each party are given by orthogonal vectors (i.e., $A_i(k_i) \perp A_i(k_i')$).
Suppose that for this specific choice of settings we define the so-called Bell polynomial \cite{werner}
\begin{align}
\label{lhvassumption}
&\mathcal{M}_n(\lambda):=\nonumber\\
&\sum_{k_1,k_2,\ldots,k_n} c(k_1,k_2,\ldots,k_n)A_1(k_1,\lambda )A_2(k_2,\lambda )\ldots A_n(k_n,\lambda).
\end{align}
It follows from $ A_i(k_i,\lambda )=\pm1$ for all $i$ that for the specific settings used 
 $-1\leq\mathcal{M}_n(\lambda) \leq 1$.
Because of linearity of the mean (i.e., the linear combination in \Eq{bellineq} can also be 
evaluated under the expectation value) the Bell inequality \Eq{bellineq} 
(for the specific choise of coefficients $c(k_1,k_2,\ldots,k_n)$) can be reformulated  as an upperbound 
 on the expectation of 
 this Bell polynomial $\mathcal{M}_n(\lambda)$.  Indeed, 
 local realism predicts the result that
\BEQ\label{avM}
|\av{\mathcal{M}_n}_\mathrm{LHV}|=|\int  \mathcal{M}_n(\lambda)\,p(\lambda)\,d \lambda|\leq 1,
\EEQ
 with the expectation value $\av{X}_{\mathrm{LHV}}:=\int_\lambda X(\lambda) p(\lambda)\,d \lambda$.
This is in fact a shorthand notation for the Bell-type inequality
\beq\label{bellineq2}
|\sum_{k_1,k_2,\ldots,k_n} c(k_1,k_2,\ldots,k_n)
E(k_1,k_2,\ldots,k_n)|
\leq 1,
\eeq
with the correlation $E(\,\cdot\,)$ given by \Eq{lr}. This the Mermin-Roy-Singh-Ardehali-Belinski\u{\i}-Klyshko inequality \cite{mkrsabk}.

 The quantum mechanical counterpart of this inequality is obtained as follows.
Choose $p(\lambda) =\delta(\lambda -\rho)$ with $\rho$ a 
quantum mechanical state of $n$ qubits, and next
substitute the measurement functionals in the Bell polynomial by the operators associated to the measurements to give its quantum counterpart which is denoted by the operator $\hat{\mathcal{M}}_n$:
 \beq
 \hat{\mathcal{M}}_n:=
\sum_{k_1,k_2,\ldots,k_n} c(k_1,k_2,\ldots,k_n)\hat{A}_1(k_1)\hat{A}_2(k_2 )\ldots \hat{A}_n(k_n).
\eeq
The local settings are anticommuting (i.e. $\{\hat{A}_i(k_i),\hat{A}_i(k_i ')\}=0$) 
since this incorporates the local orthogonality of the dichotomic observables. 

  After this conversion we obtain the well-known result \cite{mkrsabk} that quantum mechanics obeys the
inequality 
\beq\label{qmineq}
|\av{\hat{\mathcal{M}}_n}_\rho|\leq 2^{(n-1)/2},
\eeq
where the expectation value $\av{\hat{X}}_\rho:=\mathrm{Tr}[\hat{X}\rho]$. 
The upperbound can be achieved by the
maximally entangled GHZ states. 

If we set  $n=2$ in \Eq{bellineq2} we obtain the original CHSH inequality
for local realism and for $n=2$ in \Eq{qmineq} we get the Tsirelson inequality for quantum mechanics.

Chen now considers the quantum mechanical operator
\beq\label{sum}
\hat{\mathcal{V}}_n:=\hat{\mathcal{M}}_n+\hat{\mathcal{M}}_n^2,
\eeq
and shows that
\BEQ
\av{\hat{\mathcal{V}}_n}_{\rho}&:=&\mathrm{Tr}[\rho\hat{\mathcal{V}}_n]\nonumber
\\
&=&\mathrm{Tr}[\rho\hat{\mathcal{M}}_n] +
\mathrm{Tr}[\rho\hat{\mathcal{M}}_n^2]
\nonumber\\
&=&\av{\hat{\mathcal{M}}_n}_{\rho} +
\av{\hat{\mathcal{M}}_n}_{\rho}^2+\Delta(\hat{\mathcal{M}}_n)_{\rho},
\EEQ
with $\Delta(\hat{\mathcal{M}}_n)=\av{(\hat{\mathcal{M}}_n -\av{\hat{\mathcal{M}}_n}_\rho)^2}_\rho$
the variance of $\hat{\mathcal{M}}_n$ in the state $\rho$.
For separable quantum states $\rho_{\mathrm{sep}}$ he obtains the 
bound $\av{\hat{\mathcal{V}}_n}_{\rho_{\mathrm{sep}}}\leq2^{n-1}$,
which is tight since it can be achieved by a separable state 
(see \cite{footnote}).

Local realism gives \Eq{avM}, and Chen furthermore claims that
\BEQ
\Delta(\mathcal{M}_n)_\mathrm{LHV}&:=&\int  (\mathcal{M}_n(\lambda) -\av{\mathcal{M}_n}_\mathrm{LHV})^2 \, p(\lambda)\,d \lambda
\nonumber
\\
&=&\av{\mathcal{M}_n^2}_\mathrm{LHV}-\av{\mathcal{M}_n}_\mathrm{LHV}^2
\label{variance}
\\
\label{wrong}
&\leq& 1-\av{\mathcal{M}_n}_\mathrm{LHV}^2,
\EEQ
from which it follows that
\begin{align}
\av{\mathcal{V}_n}_\mathrm{LHV}&:=\av{{\mathcal{M}}_n}_{\mathrm{LHV}}+\av{{\mathcal{M}}_n^2}_{\mathrm{LHV}}\nonumber\\
&=\av{{\mathcal{M}}_n}_{\mathrm{LHV}}+\av{{\mathcal{M}}_n}_{\mathrm{LHV}}^2+
\Delta({\mathcal{M}}_n)_{\mathrm{LHV}}\leq2.
\end{align}

Chen thus obtains that
$\av{\mathcal{V}_n}_\mathrm{LHV}\leq2$ whereas 
$\av{\hat{\mathcal{V}}_n}_{\rho_{\mathrm{sep}}}\leq2^{n-1}$ and since
 this last bound can be achieved by separable states, he concludes 
 that for $n>2$ they violate the local 
 realistic inequality by an
 exponentially large factor.
 This conclusion contradicts the previous analysis
 that the correlations in separable states can be reproduced by local realism.
 So where did Chen's analysis go wrong?

\vskip0.5cm
\begin{centerline}{\bf Exposing the problematic relation}\end{centerline}
\vskip0.4cm
\noindent
Let us first note that 
in \Eq{wrong} it must have been used that
\beq\label{wrongineq}
\av{\mathcal{M}_n^2}_\mathrm{LHV}= \int (\mathcal{M}_n(\lambda))^2\,p(\lambda)\, d\lambda \leq 1,
\eeq 
which follows from $(\mathcal{M}_n(\lambda))^2\leq 1$ that on its turn follows from 
the fact that $-1\leq\mathcal{M}_n(\lambda) \leq 1$.

Using this we see that Chen's claim follows solely from the quantum mechanical inequality
\BEQ
\av{\hat{\mathcal{M}}_n^2}_\rho\leq 2^{n-1},
\label{sepM}
\EEQ
and the fact that \Eq{sepM} can be saturated for separable states. 
Comparing this to \Eq{wrongineq}  
it would seem that already for $n=2$ separable states give rise to 
correlations that can not be reproduced by local realism.

However, it is of course not $\mathcal{M}_n(\lambda)$ that is measured in any experiment, but the observables 
$A_i(k_i,\lambda)$. But this captures only part of the problem since, as we have seen 
$\mathcal{M}_n(\lambda)$ can nevertheless be used to obtain a legitimite 
Bell-type inequality (i.e.  \Eq{bellineq2}), whereas $(\mathcal{M}_n(\lambda))^2$  can not -- or so I claim. What accounts for this difference? 

A first starting point is to note that the operator
$\hat{\mathcal{V}}_n=\hat{\mathcal{M}}_n+\hat{\mathcal{M}}_n^2$  of \Eq{sum} should be 
translated into its hidden variable counterpart as 
$\mathcal{V}_n(\lambda)= \mathcal{M}_n(\lambda)+ \mathcal{M}^2_n(\lambda)$.
But in order for Chen's analysis to go through, he must have translated this into 
$\mathcal{V}_n(\lambda)= \mathcal{M}_n(\lambda)+ (\mathcal{M}_n(\lambda))^2$.
Chen must thus have assumed that 
$\mathcal{M}_n^2(\lambda)= (\mathcal{M}_n(\lambda))^2$, 
and since $\hat{\mathcal{M}}_n$ and $\hat{\mathcal{M}}_n^2$ 
commute this seems a reasonable requirement. 
Indeed, but only if the quantity 
$\mathcal{M}_n(\lambda)$ can be considered unproblematically as a hidden variable observable.  However, I will now argue that this is not the case.
 
  Let's take a closer look at the Bell polynomial $\mathcal{M}_n(\lambda)$ 
  of \Eq{lhvassumption}, and --- one is reminded of Bell's 1966 critique \cite{bell}
on von Neumann's `no-go theorem' ---  it then becomes apparent that 
the definition of
$\mathcal{M}_n(\lambda)$  uses a suspicious additivity of incompatible 
observables, since it involves different local setting ($k_i\neq k_i '$).
Because of this the Bell polynomial of \Eq{lhvassumption} can not be 
considered to be an observable that local realism determines, and is never measured as such,
 since, as Bell has taught us, 
"a measurement of a sum of noncommuting observables
cannot be made by combining trivially the results of separate observations
on the two terms -- it requires a quite distinct experiment. 
[\ldots] But this explanation of the nonadditivity of allowed values also established 
the nontriviality of the additivity of expectation values. 
The latter is quite a peculiar property of quantum mechanical states, 
not to be expected \emph{a priori}. There is no reason to demand 
it individually of the hypothetical dispersion free states 
[hidden variable states $\lambda$], whose function it is to reproduce the 
\emph{measurable} peculiarities of quantum mechanics when \emph{averaged over}.
"\cite{bell}. For Bell an expression involving noncommuting spin observables such as $[\sigma_x +\sigma_y](\lambda)$
could not be assumed to be equal to $\sigma_x(\lambda)+ \sigma_y(\lambda)$.

 If we apply Bell's lesson to Chen's analysis we obtain that 
 measurement of the Bell polynomial $\mathcal{M}_n(\lambda)$ can not be made by combining  
trivially the results of noncommuting observables.
The hidden variables $\lambda$ only determine the outcomes $A_i(k_i,\lambda)$ 
of 
individual measurements (with settings $k_i$) and
not the outcomes of measurement of the quantity $\mathcal{M}_n(\lambda)$ 
since the latter involves incompatible observables.  The only function 
of the Bell polynomial  $\mathcal{M}_n(\lambda)$  
is to allow for a shorthand notation of the Bell-type 
inequalities. Indeed, when averaged over $\lambda$  it 
gives the inequality \Eq{avM}, which by using linearity of the mean 
can be rewritten as a sum of expectation values 
in a legitimate local realistic form, namely as  the 
legitimate Bell-type inequality of \Eq{bellineq2} 
that local realism must satisfy. 
Indeed, all expectation values in the Bell-type inequality  \Eq{bellineq2}
involve only commuting (compatible) quantities, and no noncommuting ones.
So although $\mathcal{M}_n(\lambda)$ cannot be thought of as an observable specifying the sum of 
local realistic quantities determined by the hidden variable $\lambda$, and as 
such is never measured in an experiment,  when averaged over it does allow 
for a legitimate shorthand notation of the Bell-type inequality \Eq{bellineq2}.
This is the reason why the hidden variable counterpart of the operator $\hat{\mathcal{M}}_n$ 
can be safely chosen to be the Bell-polynomial 
$\mathcal{M}_n(\lambda)$. 



However, I will now argue that from a local realistic point of view 
such a manouvre cannot performed for the functional 
$(\mathcal{M}_n(\lambda))^2$. 

Firstly, $(\mathcal{M}_n(\lambda))^2$ is not 
the legitimite hidden variable counterpart of 
the operator $(\hat{\mathcal{M}}_n)^2$. Thus when averaged over it gives 
the inequality \Eq{wrongineq} but this is not a legitimite local realistic  
counterpart of the inequality \Eq{sepM}. This can be easily seen from 
the following example. Suppose we take $n=2$ and use the observables 
$A$ and $A'$ and $B$ and $B'$ for party  $1$ and $2$ respectively. Then if we 
expand $(\mathcal{M}_2(\lambda))^2$ we get 
\BEQ\label{locexp}
(\mathcal{M}_2(\lambda))^2&=& \frac{1}{4}(A^2+A'^2)(B^2 +B'^2)+
 \frac{1}{2}AA'(B^2-B'^2) \nonumber\\&&~~~~~~~~~+\, \frac{1}{2}BB'(A^2-A'^2),
\EEQ
where the dependency of A, A', B, and B' on $\lambda$ has been omitted for clarity.
However, if we expand $(\hat{\mathcal{M}}_2)^2$ (and using the local anticommutativity) we get 
\begin{align}
(\hat{\mathcal{M}}_2)^2=\frac{1}{4}(\hat{A}^2+\hat{A}'^2)\otimes(\hat{B}^2 +\hat{B}'^2)-
(\hat{A}\hat{A}'\otimes\hat{B}\hat{B}')\nonumber\\
=\frac{1}{4}(\hat{A}^2+\hat{A}'^2)\otimes(\hat{B}^2 +\hat{B}'^2)+
(\hat{A}''\otimes \hat{B}''),~~
\label{qmexp}
\end{align} 
 with 
$\hat{A}''=[\hat{A},\hat{A}']/2i$ and $\hat{B}''=[\hat{B},\hat{B}']/2i$ 
some self adjoint operators (with eigenvalues $\pm 1$) that can be thought to 
correspond to some well defined observables.
We indeed see that \Eq{qmexp} is structurally different 
from the local realistic expression \Eq{locexp}, and they can therefore not be considered to be the counterpart of eachother.  The correct 
local realistic counterpart of $(\hat{\mathcal{M}}_2)^2$ is obtained by translating \Eq{qmexp} directly into
\beq
\mathfrak{M}_2(\lambda):= \frac{1}{4}(A^2+A'^2)(B^2 +B'^2) +A''B'',
\eeq
with $A''$ and $B''$ some dichotomic $\pm1$ valued observables that are the local realistic counterpart of respectively $\hat{A}''$ and $\hat{B}''$. (The dependency of the right hand side quantitities on $\lambda$ has again been omitted for clarity).
The functional $\mathfrak{M}_2(\lambda)$ (and \emph{not} ($\mathcal{M}_2(\lambda))^2$) is the Bell-polynomial that,  when averaged over and using linearity of the mean, gives the Bell-type inequality which is the counterpart of the quantum mechanical inequality using $(\hat{\mathcal{M}}_2)^2$.
%
%

Secondly, when averaged 
over  
$(\mathcal{M}_n(\lambda))^2$ does not give a 
shorthand notation for a Bell-type inequality  
or any other constraint which local realism must obey. 
Invoking linearity of the mean does not help here.
%
%
  
The reason for this is that we end up with the above mentioned 
problem that Bell pointed out: we get expectation values 
of the products of noncommuting (incompatible) observables which 
cannot be determined by combining measurements of the individual observables. 
To see this we expand  $\mathcal{M}_n(\lambda))^2$ in 
 $\av{(\mathcal{M}_n(\lambda))^2}_{\mathrm{LHV}}$, as was done in \Eq{locexp} for $n=2$. We then get terms like 
$\av{[\ldots A_i(k_i,\lambda)A_i(k_{i}',\lambda)\ldots]}_{\mathrm{LHV}}$, which for 
$k_i \neq k_{i}'$ involve incompatible experiments that correspond to locally 
anticommuting operators in quantum mechanics.  Indeed, it is precisely this 
local noncommutativity, i.e., $\hat{A}_i(k_i)\hat{A}_i(k_i') =- 
\hat{A}_i(k_i')\hat{A}_i(k_i)$, that has no counterpart for local 
realistic observables, which is responsible for the structural differences
in \Eq{locexp} and \Eq{qmexp} and which Chen uses to get the 
exponentially diverging result that $\av{(\mathcal{M}_n(\lambda))^2}_{\mathrm{LHV}}\leq 1$ 
whereas $\av{\mathcal{\hat{M}}^2_n}_{\rho}\leq 2^{(n-1}$ (with the latter 
tight for separable quantum states). 
But we have seen that for local realism these 
expectation values 
of the products of noncommuting (incompatible) observables are problematic 
for precise the same reason as why additivity of expectation values is 
problematic for 
  the sum of noncommuting observables:
  measurement of the product
of noncommuting observables requires a quite distinct experiment
from the experiments used to measure the individual terms in the product.

Thus Bell's critique on von Neumann's no-go theorem equally well applies 
here too: "[\ldots] the formal proof does not
justify his informal conclusion"\cite{bell}, i.e., although Chen's proof is
mathematically correct and as such is interesting, his conclusion is nevertheless wanting since the 
local hidden variable theorist would not be enforced to regard the 
 assumption \Eq{wrongineq} as the legitimite counterpart of \Eq{sepM},
  nor to regard $(\mathcal{M}_n(\lambda))^2$ as 
  a legitimite shorthand notation for any sort of Bell-type inequality.  Chen's analysis thus breaks down.

It is thus not the strength of correlations or fluctuations in separable states which
ruled out local realism, but "[i]t was the arbitrary assumption of a particular (and impossible) relation
between he results of incompatible measurements either of which \emph{might} be made on a
 given occasion but only one of which can in fact be made."\cite{bell}


\vskip0.5cm
\begin{centerline}{\bf Repercussions for modern research}
 \end{centerline}\begin{centerline}
{\bf on Bell-type inequalities}
\end{centerline}
\vskip0.4cm
\noindent
In modern research on Bell-type inequalities (see \cite{mkrsabk,chen,chenco,werner}, however cf.  \cite{footnote2}) one often considers recursive definitions and shorthand notations 
in terms of Bell polynomials \cite{werner} (e.g., see \Eq{lhvassumption})
and their quantum mechanical counterparts, 
the so called Bell operators. The latter are particular 
linear combinations of operators that 
correspond to products of local observables. 
Examples of such Bell-operators are 
$\hat{\mathcal{M}}_n$ in the multi-partite setting or for example
$\hat{\mathcal{B}}=\hat{A}\hat{B}+ \hat{A}\hat{B}' +
\hat{A'}\hat{B} -\hat{A}'\hat{B}'$ for the bi-partite setting.
 In quantum mechanics these operators $\hat{\mathcal{M}}_n$ and
  $\hat{\mathcal{B}}$
can be considered to be observables themselves since a sum of self-adjoint
operators is again self-adjoint and every self-adjoint
operator is supposed to correspond to an observable.
 Furthermore, the additivity of operators gives
additivity of expectation values.
(This is the reason why the operators $\hat{\mathcal{M}}_n$ and $\hat{\mathcal{M}}_n^2$ that Chen uses can be considered to be proper quantum mechanical observables.)

Thus the so called Tsirelson inequality
expressed as  
\beq
|\av{\hat{A}\hat{B}}_\rho +\av{\hat{A}\hat{B}'}_\rho +\av{\hat{A}'\hat{B}}_\rho-\av{\hat{A}'\hat{B}'}_\rho|\leq 2\sqrt{2}
\eeq 
can thus equally well be expressed in a shorthand notation as 
$|\av{\hat{\mathcal{B}}}_\rho|\leq 2\sqrt{2}$, with $\hat{\mathcal{B}}=\hat{A}\hat{B}+ \hat{A}\hat{B}' +
\hat{A'}\hat{B} -\hat{A}'\hat{B}'$.

However, as noted by Bell and as cited before in this note,
this is additivity of expectation values is ``a quite peculiar property of QM, not to be expected
\emph{a priori}" for the hidden variable states $\lambda$.
Thus for Bell-type inequalities such a shorthand notation 
involves taking care of some crucial subtleties. 
I will now discuss three such subtleties that should be taken 
into account in deriving Bell-type inequalities, and then relate 
them to the discussion of the previous section.

(I) Firstly, the local realistic Bell polynomials are not to be regarded 
as observables. The danger is that because the operator identity in quantum mechanics
$\hat{\mathcal{B}}=\hat{A}\hat{B}+ \hat{A}\hat{B}' +
\hat{A'}\hat{B} -\hat{A}'\hat{B}'$ does indeed define a new observable $\hat{\mathcal{B}}$, one is tempted to formulate 
the hidden variable counterpart as:
\beq\label{counter}
\mathcal{B}(\lambda)= A(\lambda)B(\lambda)+A(\lambda)B'(\lambda)
+A'(\lambda)B(\lambda)-A'(\lambda)B'(\lambda).
\eeq
However, if $\mathcal{B}(\lambda)$ is not regarded as 
merely a shorthand notation for the sum of the four terms in \Eq{counter}, but
is supposed to be the counterpart of the observable
$\hat{\mathcal{B}}$, \Eq{counter}  involves the problematic additivity of eigenvalues, 
which cannot be demanded of local realism.
Indeed, four different non-compatible setups are involved
and not just one, which the notation $\mathcal{B}(\lambda)$ 
when considered as a single observable could suggest.

Thus, when deriving local hidden variable observables that depend on the hidden variable $\lambda$, only compatible experimental setups must be considered.
The difficulty of measuring incompatible observables
thus has to be explicitly taken into account in the hidden variable expression.
This is different from quantum mechanics where the incompatibility
structure already is captured in the (non-)commutativity structure of the
operators that correspond to the observables in question.

(II) Secondly, when using a shorthand notation it must be possible (by for example 
using linearity of the mean) to translate the shorthand notation into a 
legitimate Bell-type inequality which a local realist cannot but accept 
or into any other legitimate local realistic constraint. 
As an example, note that we have seen that $\mathcal{M}_n(\lambda)$ did allow for 
formulating a legitimite Bell inequality whereas 
$(\mathcal{M}_n(\lambda))^2$ did not.


(III) Thirdly, 
suppose one would indeed regard the functionals 
$\mathcal{M}_n(\lambda)$ and $\mathcal{B}(\lambda)$ to be the 
quantities of interest and regard them as observables. 
The first subtlety mentioned above shows that this is unproblematic only if they are thought of as 
being genuine irreducible observables and not to be composed out of other incompatible observables.
Then to be fair to local realism from the start
the possible values of measurement of these observables $\mathcal{M}_n(\lambda)$ and $\mathcal{B}(\lambda)$ in the local realist model should
 then be equal to the eigenvalues of the quantum mechanical counterparts
$\hat{\mathcal{M}}_n$ and $\hat{\mathcal{B}}$.  And these eigenvalues are
\{$2^{(n-1)/2},~-2^{(n-1)/2}, ~0$\} and \{$2\sqrt{2},~-2\sqrt{2},~ 0$\} respectively. The possible outcomes for the local realist quantities should equal these eigenvalues, and then no violation can be
seen to occur. Indeed, predictions for a single observation can always be mimicked by a local realistic model. 
Furthermore, if we now go back to the problematic operator $\hat{\mathcal{M}}^2_n$ that 
Chen considered, we see that it has eigenvalues \{$2^{(n-1)},~0$\}, (see footnote \cite{footnote}), 
whereas it was assumed that its local realistic counterpart has outcomes in $[-1,1]$, 
and can thus by construction, and not because of local realism, never reproduce the quantum outcomes.
 
Using these three subtleties we can now understand in a different way where Chen's analysis has gone astray:
(i) If we think of \Eq{wrongineq} as a mere shorthand notation of a complex 
summation, 
then the second subtlety is the major problem. On this reading the functional 
$(\mathcal{M}_n(\lambda))^2$ cannot be taken to give a constraint 
(i.e., \Eq{wrongineq}) which is a shorthand notation of a legitimate Bell-type 
inequality. 
(ii) However, one can argue that the analysis of Chen does not treat the Bell 
polynomial $\mathcal{M}_n(\lambda)$ in \Eq{lhvassumption} as merely a 
shorthand notation for
a complex summation. Indeed, on a different reading one can think that the 
relation of \Eq{lhvassumption} should be treated as defining a local realistic 
observable itself, since it is of this quantity that Chen considers the 
variance and the expectation value of its square. 
However, to treat  \Eq{lhvassumption} as if it was defining an observable 
makes no difference at all for quantum mechanics, but as we have seen, for 
local realism this makes a crucial difference. A local realist would either 
encounter the problem mentioned in the first subtlety (and would then face 
the Bell-type critique), or the problem mentioned in the third subtlety (and
would then not give any interesting results, i.e., local realism can then trivially 
reproduce the quantum predictions). 

We thus see that on both readings Chen's analysis breaks down.

\vskip0.5cm
\begin{centerline}{\bf Acknowledgements}
\end{centerline}
\vskip0.4cm

\noindent
I thank Jos Uffink and Dennis Dieks for very helpful comments.


\begin{thebibliography}{99}
\bibitem{hat}Note that in order to
distinghuish between the local realistic observables (i.e., measurement functionals) and the
quantum mechanical operators, I denote the quantum mechanical operators that
correspond to observables in the local realistic expression by a `hat'.
This difference will become crucial, which will become clear later on.
\bibitem{fluct}Consider a bipartite scenario. For local realism the variance $\Delta(A(\lambda)B(\lambda))$ of the joint observable $A(\lambda)B(\lambda)$ is given by $\int (A(\lambda))^2(B(\lambda))^2\,p(\lambda)\,d\lambda -(\int A(\lambda)B(\lambda)\,p(\lambda)\,d\lambda)^2$. In quantum mechanics the variance $\Delta(\hat{A}\otimes\hat{B})$ of the joint observable $\hat{A}\otimes\hat{B}$ for separable states $\rho=\sum_j p_j \rho_j^1\otimes\rho_j^2$
is given by $\sum_j p_j \mathrm{Tr}^1[\rho_j^1\hat{A}^2]\mathrm{Tr}^2[\rho_j^2\hat{B}^2]- 
(\sum_j p_j \mathrm{Tr}^1[\rho_j^1\hat{A}]\mathrm{Tr}^2[\rho_j^2\hat{B}])^2
$. This can be reproduced by the local realisic variance $\Delta(A(\lambda)B(\lambda))$ since we have that $|\mathrm{Tr}^1[\rho^1\hat{A}]|=|\av{\hat{A}}_\rho|\leq\max A(\lambda)$
(and analogously for $B$).
\bibitem{mermin} See for example the paper by Chen \cite{chen} for the explicit form of this inequality
and the choice of coefficients $c(k_1,k_2,\ldots,k_n)$ he uses. However, the explicit forms are not relevant here.
\bibitem{footnote}
The operator $\hat{\mathcal{M}}_n$  has the spectral decomposition
 $\hat{\mathcal{M}}_n= 2^{(n-1)/2}(\ket{\mathrm{GHZ}^n_+}\bra{\mathrm{GHZ}^n_+}- \ket{\mathrm{GHZ}^n_-}\bra{\mathrm{GHZ}^n_-})$, and 
the eigenvalues are (i) the values $\pm 2^{(n-1)/2}$ with
$n$-partite GHZ-states $\ket{\mathrm{GHZ}^n_\pm}=1/\sqrt{2}(\ket{0^n}\pm\ket{1^n})$ as eigenstates and (ii)  furthermore the value $0$ with degeneracy $n-2$.\\
The operator $\hat{\mathcal{M}}_n^2$  has the spectral decomposition 
$\hat{\mathcal{M}}_n^2= 2^{(n-1)}(\ket{0^n}\bra{0^n}- \ket{1^n}\bra{1^n})$, and 
eigenvalues are (i) the value $2^{(n-1)}$ (degeneracy 2)
with separable eigenstates $\ket{0^n}$
and $\ket{1^n}$ and  (ii) furthermore the value $0$ with degeneracy $n-2$.
\end{thebibliography}
\end{document}